\documentclass[aps,preprintnumbers,twocolumn,prb,amsmath,amssymb,superscriptaddress]{revtex4}

\usepackage{graphicx}
\usepackage{dcolumn}

\usepackage{bm}
\usepackage{color}

\usepackage{hyperref}

\graphicspath{{Figures/}}

\begin{document}

\title{Current-induced nucleation and dynamics of skyrmions in a Co-based Heusler alloy}
\author{W.~Akhtar}
\affiliation{Laboratoire Charles Coulomb, Universit\'{e} de Montpellier and CNRS, 34095 Montpellier, France}
\author{A. Hrabec}
\affiliation{Laboratoire Charles Coulomb, Universit\'{e} de Montpellier and CNRS, 34095 Montpellier, France}
\affiliation{Laboratoire de Physique des Solides, CNRS UMR 8502, Universit\'{e}s Paris-Sud et Paris-Saclay, 91405 Orsay Cedex, France}
\author{S.~Chouaieb}
\affiliation{Laboratoire Charles Coulomb, Universit\'{e} de Montpellier and CNRS, 34095 Montpellier, France}
\author{A. Haykal}
\affiliation{Laboratoire Charles Coulomb, Universit\'{e} de Montpellier and CNRS, 34095 Montpellier, France}
\author{I.~Gross}
\affiliation{Laboratoire Charles Coulomb, Universit\'{e} de Montpellier and CNRS, 34095 Montpellier, France}
\author{M.~Belmeguenai}
\affiliation{LSPM, CNRS-Universit\'e Paris 13, Sorbonne Paris Cit\'e, 99 avenue Jean-Baptiste Cl\'ement, 93430 Villetaneuse, France}
\author{M.~S.~Gabor}
\affiliation{Center for Superconductivity, Spintronics and Surface Science, Technical University of Cluj- Napoca, Str. Memorandumului 28, 400114 Cluj-Napoca, Romania}
\author{B.~Shields}
\affiliation{Department of Physics, University of Basel, Klingelbergstrasse 82, Basel 4056, Switzerland}
\author{P.~Maletinsky}
\affiliation{Department of Physics, University of Basel, Klingelbergstrasse 82, Basel CH-4056, Switzerland}
\author{A.~Thiaville}
\affiliation{Laboratoire de Physique des Solides, CNRS UMR 8502, Universit\'{e}s Paris-Sud et Paris-Saclay, 91405 Orsay Cedex, France}
\author{S.~Rohart}
\affiliation{Laboratoire de Physique des Solides, CNRS UMR 8502, Universit\'{e}s Paris-Sud et Paris-Saclay, 91405 Orsay Cedex, France}
\author{V.~Jacques}
\affiliation{Laboratoire Charles Coulomb, Universit\'{e} de Montpellier and CNRS, 34095 Montpellier, France}
\email{vincent.jacques@umontpellier.fr}

\begin{abstract}
We demonstrate room-temperature stabilization of dipolar magnetic skyrmions with diameters in the range of $100$~nm in a single ultrathin layer of the Heusler alloy Co$_2$FeAl (CFA) under moderate magnetic fields. Current-induced skyrmion dynamics in microwires is studied with a scanning Nitrogen-Vacancy magnetometer operating in the photoluminescence quenching mode. We first demonstrate skyrmion nucleation by spin-orbit torque and show that its efficiency can be significantly improved using tilted magnetic fields, an effect which is not specific to Heusler alloys and could be advantageous for future skyrmion-based devices. We then show that current-induced skyrmion motion remains limited by strong pinning effects, even though CFA is a magnetic material with a low magnetic damping parameter.

\end{abstract}
\date{\today}

\maketitle

\begin{figure}[t]
\includegraphics[width = 8.5 cm]{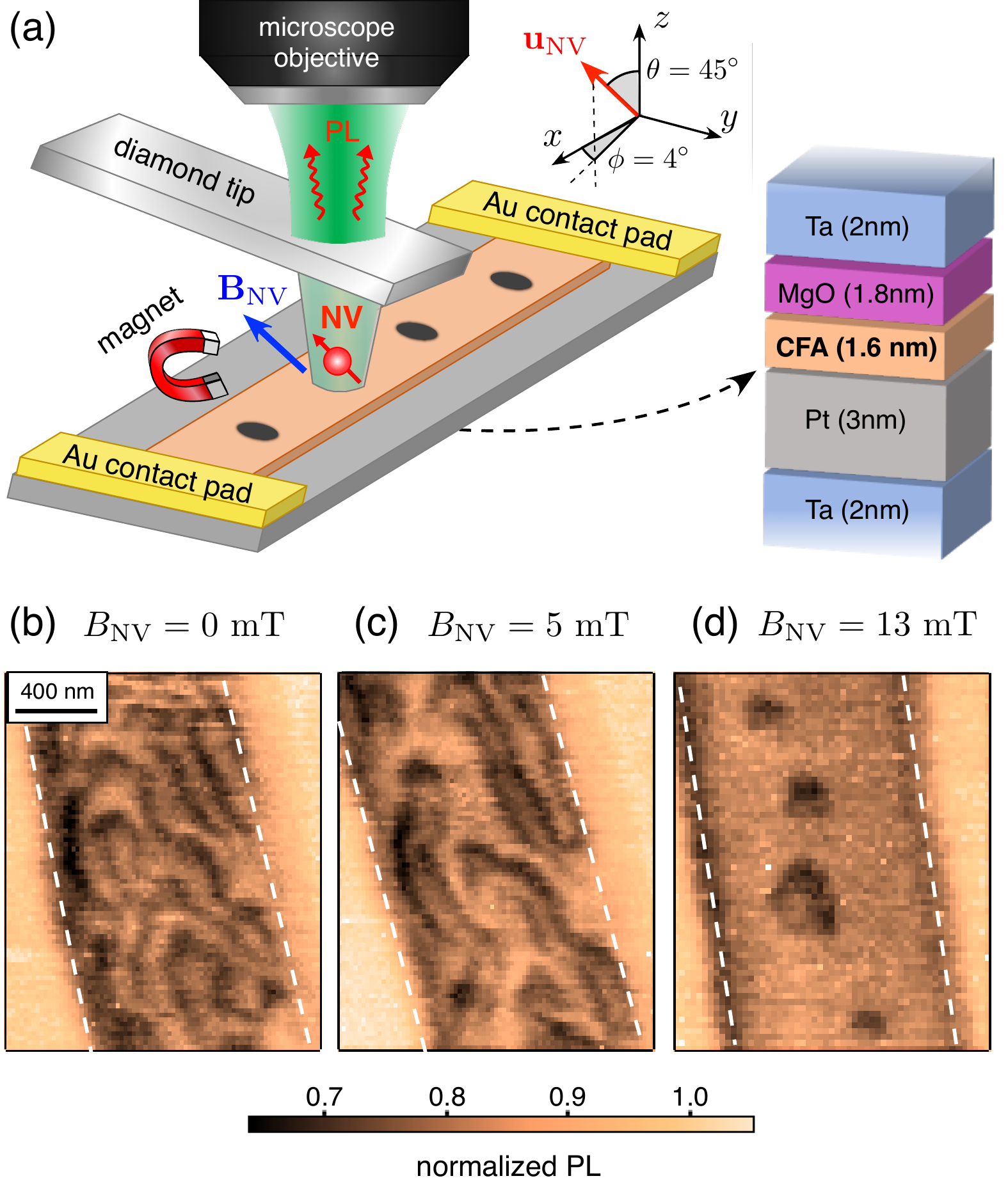}
\caption{(a) Principle of the experiment. A scanning-NV magnetometer operating in the PL quenching mode is employed to map the stray field produced above a 1-$\mu$m-wide magnetic wire made of Ta(2)$\backslash$Pt(3)$\backslash$CFA(1.6)$\backslash$MgO(1.8)$\backslash$Ta(2) (units in nm). A static magnetic field $B_{\rm NV}$ is applied along the NV axis $\bf{u_{\rm NV}}$ with a permanent magnet. The Au contact pads are used for current injection. (b-d) PL quenching images recorded for different external magnetic fields $B_{\rm NV}$ above the same area of the sample.}
\label{Fig1}
\end{figure}

Since the first proposition to use magnetic skyrmions for spintronic devices~\cite{Fert2013}, a concerted research effort has led to the experimental demonstration of several key aspects required to achieve this ambitious goal~\cite{Finocchio2016,Fert2017}. Skyrmions have now been successfully observed at room temperature in numerous types of technologically-relevant magnetic materials~\cite{Jiang2015,boulle2016,woo2016,moreau-luchaire2016,Hrabec2017,Woo2018ferri,Caretta2018}. Most of these results were obtained in magnetic multilayer stacks lacking inversion symmetry, in which the interfacial Dzyalosinskii-Moriya interaction (iDMI) promotes chiral spin textures that can be efficiently driven by spin-orbit torques~\cite{Thiaville_DMI,Sampaio2013}. In recent years, skyrmion diameters ranging from the micrometer scale down to a few tens of nanometers have been observed depending on the balance between the strength of iDMI, magnetic anisotropy and dipolar fields~\cite{buttner2018,Mantel2018}. Current-induced skyrmion motion has also been reported~\cite{woo2016,Hrabec2017,Woo2018ferri}, however with velocities remaining much smaller than those expected from seminal theoretical predictions \cite{Iwasaki2013,Sampaio2013}, which points out the key role of sample defects in skyrmion dynamics~\cite{Legrand2017,Jiang2017,Hrabec2017, gross2018}. The last key requirement of skyrmion nucleation was demonstrated by Joule heating in nanotracks \cite{Legrand2017}, current concentration in constrictions ~\cite{Jiang2015} or point contact~\cite{Hrabec2017}, and spin-orbit torque action at sample defects~\cite{buttner2017,Woo2018nucle}.\\
\indent A large variety of host materials are currently being explored with the aim of optimizing the properties of skyrmions for spintronic applications. Here, we investigate the possibility to improve current-induced skyrmion dynamics by using an Heusler alloy, a class of ferromagnetic materials with an intrinsically low magnetic damping coefficient~\cite{Mizukami2009}. Although a skyrmion phase was recently identified at low temperature in some Heusler alloys, such as Mn$_2$CoAl and Mn$_2$RhSn, by detecting a topological Hall effect~\cite{Ludbrook2017,phatak2016,rana2016}, the stabilization of magnetic skyrmions under ambient conditions has remained elusive to date and the promise of Heusler alloys for efficient skyrmion motion still needs to be explored. \\
\indent In this paper, we employ a Nitrogen-Vacancy (NV) magnetometer~\cite{Rondin2014} to demonstrate room-temperature stabilization of dipolar magnetic skyrmions with diameters in the range of $100$~nm in a single ultrathin layer of the Heusler alloy Co$_2$FeAl (CFA) under moderate magnetic fields. We then investigate skyrmion nucleation by spin-orbit torque and show that its efficiency can be significantly improved using tilted magnetic fields, an effect which is not specific to Heusler alloys and that could be used in other systems. Finally current-induced skyrmion motion is addressed, showing that pinning effects still limit skyrmion velocity, even in a host material with low magnetic damping. \\
\indent The sample used in this study is a multilayer stack deposited by magnetron sputtering on a thermally oxidized silicon wafer~\cite{Belmeguenai2018}. As shown in Figure~1(a), it consists of a CFA layer with thickness $t=1.6$~nm sandwiched between $3$~nm of Pt and $1.8$~nm of MgO. Prior to the growth of this Pt(3)$\backslash$CFA(1.6)$\backslash$MgO(1.8) heterostructure, a 2-nm-thick Ta buffer layer was deposited on the silicon substrate. A capping layer of Ta was also added to prevent oxidation of the stack. Hysteresis loops recorded with a vibrating sample magnetometer (VSM) indicate a uniaxial perpendicular magnetic anisotropy of the CFA layer $K_u=8.9\times 10^5$ J/m$^{3}$ and a saturation magnetization $M_s = 1.14\times 10^6$~A/m. In this sample, the effective anisotropy is therefore low, $K_\mathrm{eff} = 74$~kJ/m$^3$, which induces a low domain wall energy and therefore favors skyrmion stabilization through demagnetizing effects~\cite{buttner2018,Mantel2018}. The strength of iDMI, which is another key parameter for stabilizing skyrmions, was determined by monitoring the non-reciprocal propagation of spin waves with Brillouin light spectroscopy (BLS) in the Damon-Eshbach geometry~\cite{KaiPRL2015,Belmeguenai2015}, leading to an effective iDMI constant $D=-0.70\pm 0.06$~mJ/m$^2$, {\it i.e.} with the same sign as Pt$\backslash$Co$\backslash$AlO$_{x}$. Finally, a damping parameter $\alpha\sim 0.02$ was inferred from BLS measurements. The multilayer stack was patterned into $1$-$\mu­$m-wide magnetic wires by using a combination of e-beam lithography and ion milling. Ti$\backslash$Au electrical contact pads were then deposited for applying current pulses into the magnetic wires.\\
\indent Magnetic imaging was performed with a scanning-NV magnetometer operating in the photoluminescence (PL) quenching mode under ambient conditions~\cite{Rondin2014}. As sketched in Fig.1(a), a single NV defect located at the apex of a diamond tip~\cite{Maletinsky2012,Appel2016} is integrated into an atomic force microscope (AFM) and scanned above the magnetic sample. The AFM system is combined with a confocal optical microscope that collects the NV defect PL signal at each point of the scan.  When a magnetic field larger than $\sim 5$~mT is applied {\it perpendicular} to the NV defect quantization axis, an efficient mixing of its electron spin sub-levels leads to an overall reduction of the PL intensity~\cite{Epstein2005,Lai2009}. Here, this magnetic-field-induced PL quenching is exploited to map regions of the sample producing large stray fields with a spatial resolution fixed by the NV-to-sample distance $d_{\rm NV}$~\cite{Rondin2012,Tetienne2012}. This parameter was measured through the calibration procedure described in Ref.~[\onlinecite{Hingant2015}], leading to $d_{\rm NV}\sim80$~nm. Figure 1(b) shows a typical PL quenching image recorded at zero field above a $1$-$\mu{\rm m}$-wide wire of the Pt$\backslash$CFA$\backslash$MgO stack. Dark PL contours indicate high stray field areas, which correspond to a network of domain walls organized in a worm-like structure. Compared to magnetic force microscopy, the main advantage of scanning-NV magnetometry is here the absence of magnetic back-action on the sample. This is particularly important for the study of spin textures in ultrathin films, which are often highly sensitive to magnetic perturbations.

Nucleation of isolated dipolar skyrmions was first achieved by applying a static magnetic field. In order to avoid parasitic PL quenching effects, the external field must be applied along the NV defect quantization axis $\mathbf{u}_{\rm NV}$, which is characterized by the independently measured polar angles $(\theta\approx45^{\circ},\phi\approx4^{\circ})$ in the $(x,y,z)$ sample reference frame [see Fig. 1(a)]. For a magnetic field $B_{\rm NV}=5$~mT applied along the NV axis, the magnetic domains start to shrink, and for $B_{\rm NV}=13$~mT the worm-like domain structure fully collapses into isolated skyrmions, with a minimum diameter in the range of $100$~nm. The variation in shape and size of the resulting skyrmions is linked to pinning effects arising from structural disorder in the sample~\cite{Zeissler2017,Juge2018,gross2018}.

\begin{figure}[t]
\includegraphics[width = 7.3cm]{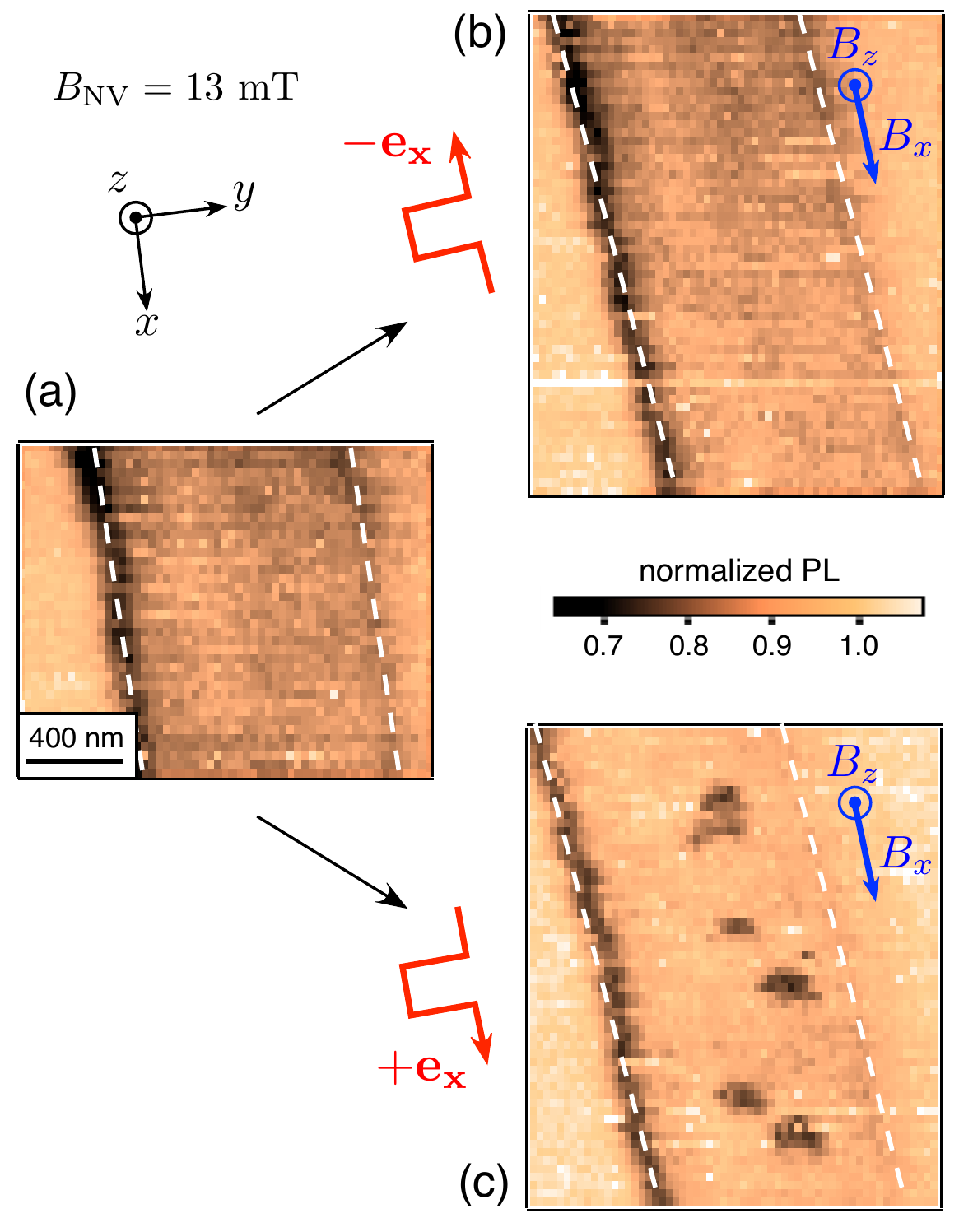}
\caption{Current-induced skyrmion nucleation. (a) PL quenching image recorded over a saturated magnetic wire. The same wire is imaged after applying a bunch of three 30-ns current pulses with $J= 6 \times 10^{11}$~A/m$^2$ (b) along the $-\mathbf{e}_x$ direction and (c) along the $+\mathbf{e}_x$ direction. These experiments are performed with an external magnetic field $B_{\rm NV}=13$~mT, which correspond to an in-plane field component $B_x\approx 9$~mT (see blue arrows).}
\label{Fig1}
\end{figure}

We now investigate skyrmion nucleation induced by short current pulses injected into the magnetic wire. As shown in Fig. 2(a), the experiments start with a fully saturated wire and a magnetic field $B_{\rm NV} =13$~mT applied along the NV quantization axis. We stress that this corresponds to an in-plane field component $B_x\approx 9$~mT along the $+\mathbf{e}_x$ direction. A bunch of three 30-ns pulses with a current density $J=6\times  10^{11}$~A/m$^2$ was first applied along the $-\mathbf{e}_x$ direction. In these conditions, skyrmions are not nucleated in the magnetic wire [Fig.~2(b)]. By simply reversing the current direction to $+\mathbf{e}_x$, nucleation becomes highly efficient [Fig.~2(c)]. Such an asymmetric nucleation process with the current polarity cannot be explained solely by a thermal activation through Joule heating~\cite{Legrand2017}. The main driving force of the nucleation process is here attributed to the spin orbit torque resulting from the spin Hall effect (SHE) in the Pt layer, as explained below.

The SHE leads to a spin accumulation at the Pt$\backslash$CFA interface, which results in a damping-like torque associated with an effective magnetic field~\cite{loconte2014} $$\mathbf{H_{\rm SHE}} = \frac{\hbar\theta_H}{2e\mu_0M_st}(\mathbf{e_z}\times \mathbf{J})\times \mathbf{m} \ ,$$ where $\theta_H$ denotes the spin Hall angle, which is positive for Pt. The perpendicular magnetization component $m_z$ leads to a SHE field along $\pm \mathbf{e}_x$, with a sign fixed by the current direction. This effective field induces a slight tilt of the magnetization but does not favor magnetization reversal for a particular current polarity~\cite{liu2012}. However, in our experiments the external magnetic field $B_x$ breaks the symmetry by producing an in-plane magnetization component $m_x$, which modifies the effect of the SHE-induced torque and promotes a switching directionality, as previously invoked in the context of magnetization reversal in thin magnetic nanodots~\cite{liu2012}. More precisely, the $m_x$ component leads to an effective SHE field perpendicular to the sample surface, which can switch the magnetization and assist skyrmion nucleation. In our case, the in-plane component of the external field being along $+\mathbf{e}_x$, $m_x$ is positive, and $H_{\rm SHE}$ has a component along $-\mathbf{e}_z$ (resp. $+\mathbf{e}_z$) when the current is applied along $+\mathbf{e}_x$ (resp. $-\mathbf{e}_x$). Starting from a saturated state with $m_z>0$, skyrmion nucleation is therefore expected to be more efficient for a current applied along $+\mathbf{e}_x$, in agreement with the experimental observation.

\begin{figure}[b]
\includegraphics[width = 7.9cm]{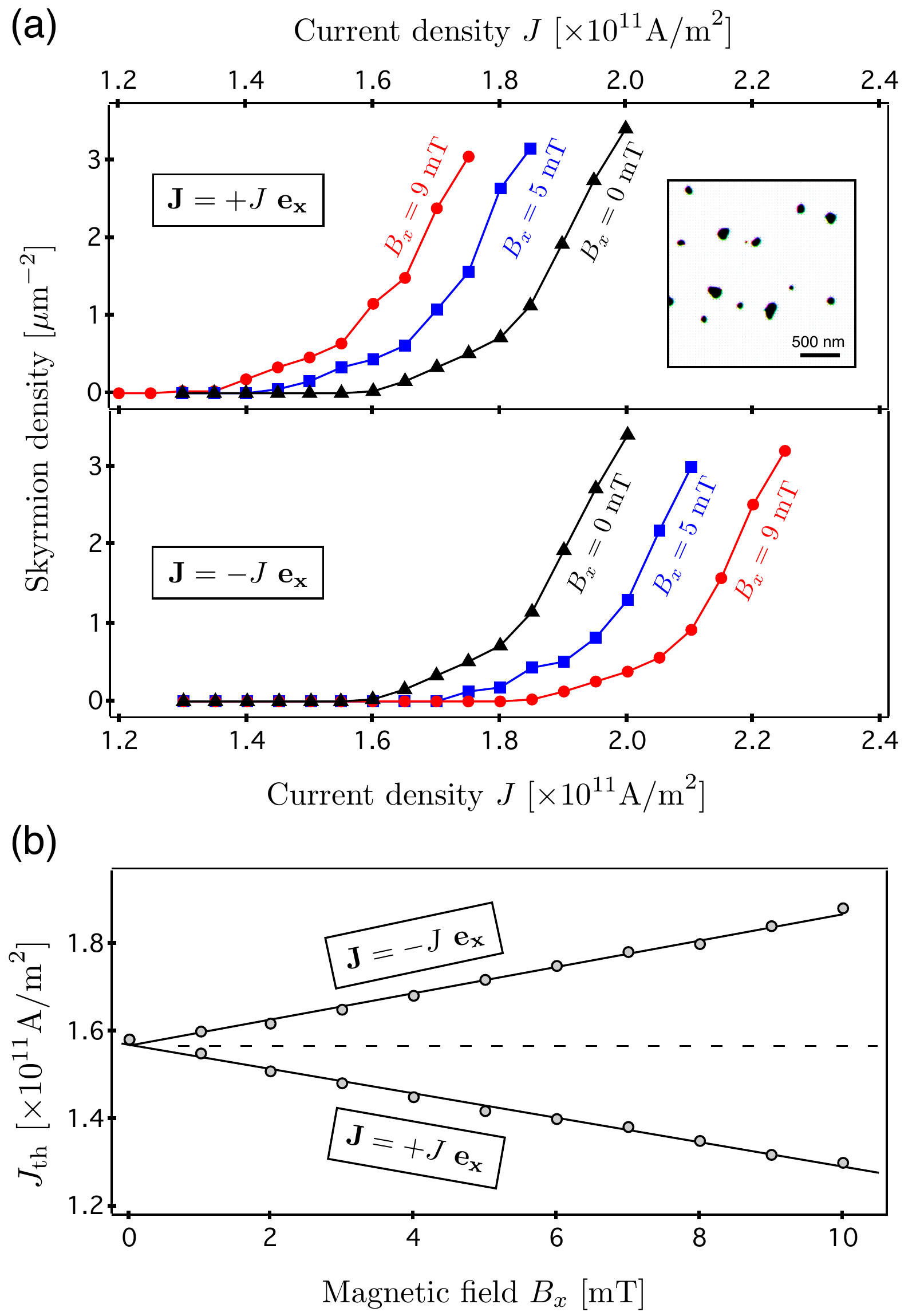}
\caption{Simulations of the skyrmion nucleation process. (a) Variation of the nucleated skyrmion density as a function of the current density and polarity, for three values of the in-plane magnetic field $B_x$. The inset shows a typical simulated magnetization map obtained for  $B_x=9$~mT and $J= 1.65\times10^{11}$~A/m$^2$ applied along $\mathbf{+e_x}$. (b) Current threshold for skyrmion nucleation, $J_{\rm th}$, as a function of $B_x$ for the two different current polarities.}
\label{simulation_figure}
\end{figure}

This qualitative interpretation was confirmed by micromagnetic simulations of the nucleation process using the MuMax3 code~\cite{MuMax} with magnetic parameters extracted from independent measurements. Thickness fluctuations of the magnetic layer with a relative amplitude of 2~\% were included in the simulation following the procedure described in Ref.~[\onlinecite{gross2018}]. Such a structural disorder provides a network of nucleation centers in the sample. A magnetic field $B_z=9$~mT was always applied perpendicular to the sample surface and the nucleation process was simulated by applying a single 2-ns long current pulse. The initial state was chosen to be uniform along the perpendicular field direction, so that nucleated skyrmion cores are opposite to this field. The skyrmion density was extracted from the simulated magnetization maps, while varying (i) the current density and (ii) the in-plane magnetic field component~$B_x$. We note that these simulations do not intend to quantitatively reproduce the experimental observations because we lack an accurate description of several key parameters of the experiment, such as the pinning landscape induced by structural disorder or thermal activation.

Simulations of the nucleated skyrmion density as a function of the current density are shown in Fig.~3(a). Without in-plane magnetic field, skyrmion nucleation is obtained above a current threshold $J_{\rm th}\approx 1.6\times10^{11}$~A/m$^2$, independently of the current polarity. This nucleation threshold is found to be tuned by an in-plane magnetic field. More precisely, $J_{\rm th}$ decreases when the current is applied along the in-plane field direction ($\mathbf{+e_x}$), and increases when the current direction is reversed. As discussed earlier, this effect is expected from the symmetry of the SHE-induced torque. This is further illustrated by Figure~3(b) showing the evolution of $J_{\rm th}$ as a function of the in-plane magnetic field. For $B_x=10$~mT, the nucleation threshold decreases (resp. increases) by $\approx 20\%$ when the current is applied along $\mathbf{+e_x}$ (resp. $\mathbf{-e_x}$). This effect leads to an asymmetric nucleation efficiency with the current polarity, as observed in our experiments. Although recent reports in the literature have also invoked SHE-induced torque as a driving force to explain deterministic skyrmion nucleation~\cite{buttner2017,Woo2018nucle}, our work demonstrates that its efficiency can be significantly improved by applying an additional in-plane magnetic field.

\begin{figure}[t]
\includegraphics[width = 8.7cm]{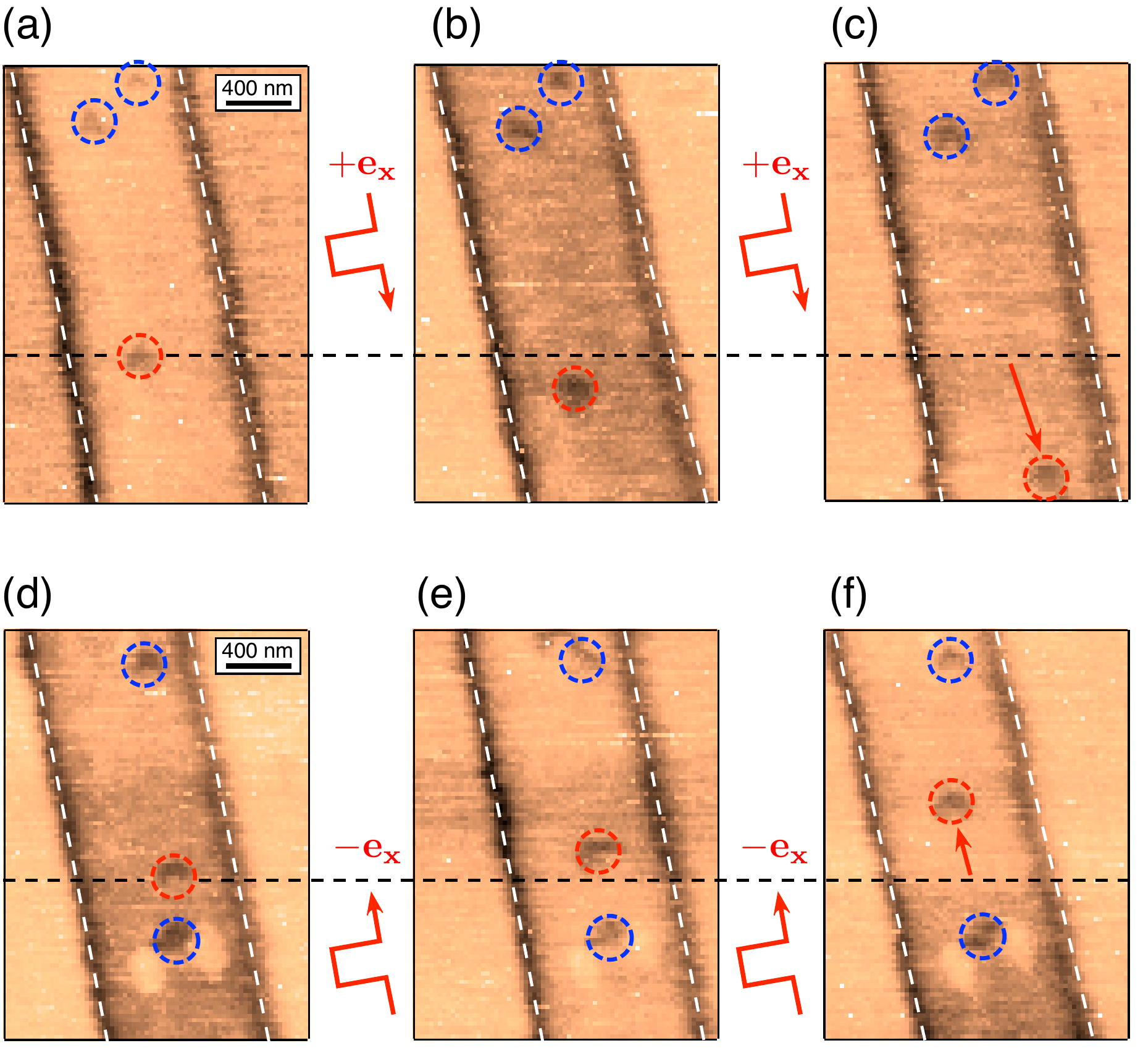}
\caption{ Current-induced skyrmion motion. (a-c) Series of PL quenching images recorded above a magnetic wire after applying subsequent 30-ns pulses with a current density $J=3\times10^{11}$~A/m$^2$ along the $\mathbf{+e_x}$ direction. (d-e) Same experiments performed on another wire while reversing the current polarity and using 15-ns pulses with $J = 4.5\times 10^{11}$~A/m$^2$. All images are recorded with a magnetic field $B_{\rm NV} =13$~mT applied along the NV axis.}
\label{Fig4}
\end{figure}

Once the microwires are filled with isolated skyrmions, we then investigate their current-induced motion. To this end, PL quenching images were recorded after applying subsequent single current pulses with a magnetic field $B_{\rm NV} =13$~mT applied along the NV axis. In order to avoid nucleation or annihilation of skyrmions by Joule heating and SHE-induced torque, the current density was decreased to study their dynamics. Figure~4(a-c) shows the initial magnetic configuration of a wire and those observed after applying single 30-ns current pulses along the $\mathbf{+e_x}$ direction with a current density $J=3\times10^{11}$~A/m$^2$. Although most of the magnetic textures remain pinned in the wire, current-induced motion can be observed for some skyrmions. Importantly, the motion is found against the electron flow. This observation was confirmed in another wire in which the current polarity was reversed, as shown in Fig.~4(d-f). This suggests that the SHE is the principal source of motion and that the skyrmions are of N\'eel type with a left-handed chirality ~\cite{Emori2013,Sampaio2013}, in agreement with the negative value of the iDMI constant measured by Brillouin light spectroscopy. We note that skyrmion displacements are not equal for successive current pulses owing to pinning effects. The motion is thus in the hopping regime, which prevents a clear observation of the skyrmion Hall effect~\cite{Jiang2016}. Nevertheless, based on the data shown in Fig.~4, we can estimate a velocity in the range of $\approx 10$~m/s. It is worth mentioning that motion at higher current density could not be studied owing to parasitic skyrmion nucleation or annihilation effects. Although thin layers of CFA exhibit a low damping parameter, our observations indicate that skyrmion dynamics remain limited by strong pinning effects in this material. 

In conclusion, we have used scanning-NV magnetometry in photoluminescence quenching mode to image current-induced nucleation and dynamics of dipolar skyrmions in a Co-based Heusler alloy at room temperature. Skyrmion nucleation was found to be asymmetric with the current polarity. This asymmetry is attributed to the SHE-induced spin orbit torque in the presence of an in-plane magnetic field, which is further corroborated by micromagnetic simulations. This effect suggests a path towards efficient, deterministic generation of skyrmions via spin-orbit torques, which will be advantageous for future skyrmion-based devices. Current-induced skyrmion motion in our experiments remains limited by strong pinning effect, pointing out the key role of sample defects in skyrmion dynamics even in materials with intrinsically low magnetic damping. \\

\noindent {\it Acknowledgements:} This research has been supported by the European Research Council (ERC-StG-2014, {\sc Imagine}), the French Agence Nationale de la Recherche (ANR-17-CE24-0025, {\sc Topsky}) and the DARPA TEE program, through grant MIPR\#HR0011831554. B.~M. acknowledges financial support of the Conseil r\'egional Ile-de-France through the DIM NanoK (BIDUL project). M. S. G. acknowledges financial support of MRI - UEFISCDI through PN-III-P1-1.1-TE-2016-2131-SOTMEM grant number no. 24/02.05.2018. P. M. acknowledges financial support from the Swiss National Science Foundation Grant No. 169321.

\bibliography{CFA}

\providecommand{\noopsort}[1]{}\providecommand{\singleletter}[1]{#1}%
\begin{thebibliography}{42}
\expandafter\ifx\csname natexlab\endcsname\relax\def\natexlab#1{#1}\fi
\expandafter\ifx\csname bibnamefont\endcsname\relax
  \def\bibnamefont#1{#1}\fi
\expandafter\ifx\csname bibfnamefont\endcsname\relax
  \def\bibfnamefont#1{#1}\fi
\expandafter\ifx\csname citenamefont\endcsname\relax
  \def\citenamefont#1{#1}\fi
\expandafter\ifx\csname url\endcsname\relax
  \def\url#1{\texttt{#1}}\fi
\expandafter\ifx\csname urlprefix\endcsname\relax\def\urlprefix{URL }\fi
\providecommand{\bibinfo}[2]{#2}
\providecommand{\eprint}[2][]{\url{#2}}

\bibitem[{\citenamefont{Fert et~al.}(2013)\citenamefont{Fert, Cros, and
  Sampaio}}]{Fert2013}
\bibinfo{author}{\bibfnamefont{A.}~\bibnamefont{Fert}},
  \bibinfo{author}{\bibfnamefont{V.}~\bibnamefont{Cros}}, \bibnamefont{and}
  \bibinfo{author}{\bibfnamefont{J.}~\bibnamefont{Sampaio}},
  \bibinfo{journal}{Nat. Nanotech.} \textbf{\bibinfo{volume}{8}},
  \bibinfo{pages}{152} (\bibinfo{year}{2013}).

\bibitem[{\citenamefont{Finocchio et~al.}(2016)\citenamefont{Finocchio,
  B\"uttner, Tomasello, Carpentieri, and Kl\"aui}}]{Finocchio2016}
\bibinfo{author}{\bibfnamefont{G.}~\bibnamefont{Finocchio}},
  \bibinfo{author}{\bibfnamefont{F.}~\bibnamefont{B\"uttner}},
  \bibinfo{author}{\bibfnamefont{R.}~\bibnamefont{Tomasello}},
  \bibinfo{author}{\bibfnamefont{M.}~\bibnamefont{Carpentieri}},
  \bibnamefont{and} \bibinfo{author}{\bibfnamefont{M.}~\bibnamefont{Kl\"aui}},
  \bibinfo{journal}{J. Phys. D: Appl. Phys.} \textbf{\bibinfo{volume}{49}},
  \bibinfo{pages}{423001} (\bibinfo{year}{2016}).

\bibitem[{\citenamefont{Fert et~al.}(2017)\citenamefont{Fert, Reyren, and
  Cros}}]{Fert2017}
\bibinfo{author}{\bibfnamefont{A.}~\bibnamefont{Fert}},
  \bibinfo{author}{\bibfnamefont{N.}~\bibnamefont{Reyren}}, \bibnamefont{and}
  \bibinfo{author}{\bibfnamefont{V.}~\bibnamefont{Cros}},
  \bibinfo{journal}{Nature Reviews Materials} \textbf{\bibinfo{volume}{2}},
  \bibinfo{pages}{17031} (\bibinfo{year}{2017}).

\bibitem[{\citenamefont{Jiang et~al.}(2015)\citenamefont{Jiang, Upadhyaya,
  Zhang, Yu, Jungfleisch, Fradin, Pearson, Tserkovnyak, Wang, Heinonen
  et~al.}}]{Jiang2015}
\bibinfo{author}{\bibfnamefont{W.}~\bibnamefont{Jiang}},
  \bibinfo{author}{\bibfnamefont{P.}~\bibnamefont{Upadhyaya}},
  \bibinfo{author}{\bibfnamefont{W.}~\bibnamefont{Zhang}},
  \bibinfo{author}{\bibfnamefont{G.}~\bibnamefont{Yu}},
  \bibinfo{author}{\bibfnamefont{M.~B.} \bibnamefont{Jungfleisch}},
  \bibinfo{author}{\bibfnamefont{F.~Y.} \bibnamefont{Fradin}},
  \bibinfo{author}{\bibfnamefont{J.~E.} \bibnamefont{Pearson}},
  \bibinfo{author}{\bibfnamefont{Y.}~\bibnamefont{Tserkovnyak}},
  \bibinfo{author}{\bibfnamefont{K.~L.} \bibnamefont{Wang}},
  \bibinfo{author}{\bibfnamefont{O.}~\bibnamefont{Heinonen}},
  \bibnamefont{et~al.}, \bibinfo{journal}{Science}
  \textbf{\bibinfo{volume}{349}}, \bibinfo{pages}{283} (\bibinfo{year}{2015}).

\bibitem[{\citenamefont{Boulle et~al.}(2016)\citenamefont{Boulle, Vogel, Yang,
  Pizzini, de~Souza~Chaves, Locatelli, Mentes, Sala, Buda-Prejbeanu, Klein
  et~al.}}]{boulle2016}
\bibinfo{author}{\bibfnamefont{O.}~\bibnamefont{Boulle}},
  \bibinfo{author}{\bibfnamefont{J.}~\bibnamefont{Vogel}},
  \bibinfo{author}{\bibfnamefont{H.}~\bibnamefont{Yang}},
  \bibinfo{author}{\bibfnamefont{S.}~\bibnamefont{Pizzini}},
  \bibinfo{author}{\bibfnamefont{D.}~\bibnamefont{de~Souza~Chaves}},
  \bibinfo{author}{\bibfnamefont{A.}~\bibnamefont{Locatelli}},
  \bibinfo{author}{\bibfnamefont{T.~O.} \bibnamefont{Mentes}},
  \bibinfo{author}{\bibfnamefont{A.}~\bibnamefont{Sala}},
  \bibinfo{author}{\bibfnamefont{L.~D.} \bibnamefont{Buda-Prejbeanu}},
  \bibinfo{author}{\bibfnamefont{O.}~\bibnamefont{Klein}},
  \bibnamefont{et~al.}, \bibinfo{journal}{Nat. Nanotech.}
  \textbf{\bibinfo{volume}{11}}, \bibinfo{pages}{449} (\bibinfo{year}{2016}).

\bibitem[{\citenamefont{Woo et~al.}(2016)\citenamefont{Woo, Litzius,
  Kr\"{u}ger, Im, Caretta, Richter, Mann, Krone, Reeve, Weigand
  et~al.}}]{woo2016}
\bibinfo{author}{\bibfnamefont{S.}~\bibnamefont{Woo}},
  \bibinfo{author}{\bibfnamefont{K.}~\bibnamefont{Litzius}},
  \bibinfo{author}{\bibfnamefont{B.}~\bibnamefont{Kr\"{u}ger}},
  \bibinfo{author}{\bibfnamefont{M.-Y.} \bibnamefont{Im}},
  \bibinfo{author}{\bibfnamefont{L.}~\bibnamefont{Caretta}},
  \bibinfo{author}{\bibfnamefont{K.}~\bibnamefont{Richter}},
  \bibinfo{author}{\bibfnamefont{M.}~\bibnamefont{Mann}},
  \bibinfo{author}{\bibfnamefont{A.}~\bibnamefont{Krone}},
  \bibinfo{author}{\bibfnamefont{R.~M.} \bibnamefont{Reeve}},
  \bibinfo{author}{\bibfnamefont{M.}~\bibnamefont{Weigand}},
  \bibnamefont{et~al.}, \bibinfo{journal}{Nat. Mater.}
  \textbf{\bibinfo{volume}{15}}, \bibinfo{pages}{501} (\bibinfo{year}{2016}).

\bibitem[{\citenamefont{Moreau-Luchaire
  et~al.}(2016)\citenamefont{Moreau-Luchaire, Moutafis, Reyren, Sampaio, Vaz,
  Van~Horne, Bouzehouane, Garcia, Deranlot, Warnicke
  et~al.}}]{moreau-luchaire2016}
\bibinfo{author}{\bibfnamefont{C.}~\bibnamefont{Moreau-Luchaire}},
  \bibinfo{author}{\bibfnamefont{C.}~\bibnamefont{Moutafis}},
  \bibinfo{author}{\bibfnamefont{N.}~\bibnamefont{Reyren}},
  \bibinfo{author}{\bibfnamefont{J.}~\bibnamefont{Sampaio}},
  \bibinfo{author}{\bibfnamefont{C.~A.~F.} \bibnamefont{Vaz}},
  \bibinfo{author}{\bibfnamefont{N.}~\bibnamefont{Van~Horne}},
  \bibinfo{author}{\bibfnamefont{K.}~\bibnamefont{Bouzehouane}},
  \bibinfo{author}{\bibfnamefont{K.}~\bibnamefont{Garcia}},
  \bibinfo{author}{\bibfnamefont{C.}~\bibnamefont{Deranlot}},
  \bibinfo{author}{\bibfnamefont{P.}~\bibnamefont{Warnicke}},
  \bibnamefont{et~al.}, \bibinfo{journal}{Nat. Nanotech.}
  \textbf{\bibinfo{volume}{11}}, \bibinfo{pages}{444} (\bibinfo{year}{2016}).

\bibitem[{\citenamefont{Hrabec et~al.}(2017)\citenamefont{Hrabec, Sampaio,
  Belmeguenai, Gross, Weil, Ch\'erif, Stashkevich, Jacques, Thiaville, and
  Rohart}}]{Hrabec2017}
\bibinfo{author}{\bibfnamefont{A.}~\bibnamefont{Hrabec}},
  \bibinfo{author}{\bibfnamefont{J.}~\bibnamefont{Sampaio}},
  \bibinfo{author}{\bibfnamefont{M.}~\bibnamefont{Belmeguenai}},
  \bibinfo{author}{\bibfnamefont{I.}~\bibnamefont{Gross}},
  \bibinfo{author}{\bibfnamefont{R.}~\bibnamefont{Weil}},
  \bibinfo{author}{\bibfnamefont{S.~M.} \bibnamefont{Ch\'erif}},
  \bibinfo{author}{\bibfnamefont{A.}~\bibnamefont{Stashkevich}},
  \bibinfo{author}{\bibfnamefont{V.}~\bibnamefont{Jacques}},
  \bibinfo{author}{\bibfnamefont{A.}~\bibnamefont{Thiaville}},
  \bibnamefont{and} \bibinfo{author}{\bibfnamefont{S.}~\bibnamefont{Rohart}},
  \bibinfo{journal}{Nat. Commun.} \textbf{\bibinfo{volume}{8}},
  \bibinfo{pages}{15765} (\bibinfo{year}{2017}).

\bibitem[{\citenamefont{Woo et~al.}(2018{\natexlab{a}})\citenamefont{Woo, Song,
  Zhang, Zhou, Ezawa, Liu, Finizio, Raabe, Lee, Kim et~al.}}]{Woo2018ferri}
\bibinfo{author}{\bibfnamefont{S.}~\bibnamefont{Woo}},
  \bibinfo{author}{\bibfnamefont{K.~M.} \bibnamefont{Song}},
  \bibinfo{author}{\bibfnamefont{X.}~\bibnamefont{Zhang}},
  \bibinfo{author}{\bibfnamefont{Y.}~\bibnamefont{Zhou}},
  \bibinfo{author}{\bibfnamefont{M.}~\bibnamefont{Ezawa}},
  \bibinfo{author}{\bibfnamefont{X.}~\bibnamefont{Liu}},
  \bibinfo{author}{\bibfnamefont{S.}~\bibnamefont{Finizio}},
  \bibinfo{author}{\bibfnamefont{J.}~\bibnamefont{Raabe}},
  \bibinfo{author}{\bibfnamefont{N.~J.} \bibnamefont{Lee}},
  \bibinfo{author}{\bibfnamefont{S.-I.} \bibnamefont{Kim}},
  \bibnamefont{et~al.}, \bibinfo{journal}{Nat. Commun.}
  \textbf{\bibinfo{volume}{9}}, \bibinfo{pages}{959}
  (\bibinfo{year}{2018}{\natexlab{a}}).

\bibitem[{\citenamefont{Caretta et~al.}(2018)\citenamefont{Caretta, Mann,
  B\"{u}ttner, Ueda, Pfau, G\"{u}nther, Hessing, Churikova, Klose, Schneider
  et~al.}}]{Caretta2018}
\bibinfo{author}{\bibfnamefont{L.}~\bibnamefont{Caretta}},
  \bibinfo{author}{\bibfnamefont{M.}~\bibnamefont{Mann}},
  \bibinfo{author}{\bibfnamefont{F.}~\bibnamefont{B\"{u}ttner}},
  \bibinfo{author}{\bibfnamefont{K.}~\bibnamefont{Ueda}},
  \bibinfo{author}{\bibfnamefont{B.}~\bibnamefont{Pfau}},
  \bibinfo{author}{\bibfnamefont{C.~M.} \bibnamefont{G\"{u}nther}},
  \bibinfo{author}{\bibfnamefont{P.}~\bibnamefont{Hessing}},
  \bibinfo{author}{\bibfnamefont{A.}~\bibnamefont{Churikova}},
  \bibinfo{author}{\bibfnamefont{C.}~\bibnamefont{Klose}},
  \bibinfo{author}{\bibfnamefont{M.}~\bibnamefont{Schneider}},
  \bibnamefont{et~al.}, \bibinfo{journal}{Nat. Nanotech.}
  (\bibinfo{year}{2018}).

\bibitem[{\citenamefont{Thiaville et~al.}(2012)\citenamefont{Thiaville, Rohart,
  Ju\'{e}, Cros, and Fert}}]{Thiaville_DMI}
\bibinfo{author}{\bibfnamefont{A.}~\bibnamefont{Thiaville}},
  \bibinfo{author}{\bibfnamefont{S.}~\bibnamefont{Rohart}},
  \bibinfo{author}{\bibfnamefont{E.}~\bibnamefont{Ju\'{e}}},
  \bibinfo{author}{\bibfnamefont{V.}~\bibnamefont{Cros}}, \bibnamefont{and}
  \bibinfo{author}{\bibfnamefont{A.}~\bibnamefont{Fert}},
  \bibinfo{journal}{Europhys. Lett.} \textbf{\bibinfo{volume}{100}},
  \bibinfo{eid}{57002} (\bibinfo{year}{2012}).

\bibitem[{\citenamefont{Sampaio et~al.}(2013)\citenamefont{Sampaio, Cros,
  Rohart, Thiaville, and Fert}}]{Sampaio2013}
\bibinfo{author}{\bibfnamefont{J.}~\bibnamefont{Sampaio}},
  \bibinfo{author}{\bibfnamefont{V.}~\bibnamefont{Cros}},
  \bibinfo{author}{\bibfnamefont{S.}~\bibnamefont{Rohart}},
  \bibinfo{author}{\bibfnamefont{A.}~\bibnamefont{Thiaville}},
  \bibnamefont{and} \bibinfo{author}{\bibfnamefont{A.}~\bibnamefont{Fert}},
  \bibinfo{journal}{Nat. Nanotech.} \textbf{\bibinfo{volume}{8}},
  \bibinfo{pages}{839} (\bibinfo{year}{2013}).

\bibitem[{\citenamefont{B\"uttner et~al.}(2018)\citenamefont{B\"uttner, Lemesh,
  and Beach}}]{buttner2018}
\bibinfo{author}{\bibfnamefont{F.}~\bibnamefont{B\"uttner}},
  \bibinfo{author}{\bibfnamefont{I.}~\bibnamefont{Lemesh}}, \bibnamefont{and}
  \bibinfo{author}{\bibfnamefont{G.}~\bibnamefont{Beach}},
  \bibinfo{journal}{Sci. Rep.} \textbf{\bibinfo{volume}{8}},
  \bibinfo{pages}{4464} (\bibinfo{year}{2018}).

\bibitem[{\citenamefont{Bernand-Mantel
  et~al.}(2018)\citenamefont{Bernand-Mantel, Camosi, Wartelle, Rougemaille,
  Darques, and Ranno}}]{Mantel2018}
\bibinfo{author}{\bibfnamefont{A.}~\bibnamefont{Bernand-Mantel}},
  \bibinfo{author}{\bibfnamefont{L.}~\bibnamefont{Camosi}},
  \bibinfo{author}{\bibfnamefont{A.}~\bibnamefont{Wartelle}},
  \bibinfo{author}{\bibfnamefont{N.}~\bibnamefont{Rougemaille}},
  \bibinfo{author}{\bibfnamefont{M.}~\bibnamefont{Darques}}, \bibnamefont{and}
  \bibinfo{author}{\bibfnamefont{L.}~\bibnamefont{Ranno}},
  \bibinfo{journal}{SciPost Phys.} \textbf{\bibinfo{volume}{4}},
  \bibinfo{pages}{27} (\bibinfo{year}{2018}).

\bibitem[{\citenamefont{Iwasaki et~al.}(2013)\citenamefont{Iwasaki, Mochizuki,
  and Nagaosa}}]{Iwasaki2013}
\bibinfo{author}{\bibfnamefont{J.}~\bibnamefont{Iwasaki}},
  \bibinfo{author}{\bibfnamefont{M.}~\bibnamefont{Mochizuki}},
  \bibnamefont{and} \bibinfo{author}{\bibfnamefont{N.}~\bibnamefont{Nagaosa}},
  \bibinfo{journal}{Nat. Commun.} \textbf{\bibinfo{volume}{4}},
  \bibinfo{pages}{1463} (\bibinfo{year}{2013}).

\bibitem[{\citenamefont{Legrand et~al.}(2017)\citenamefont{Legrand,
  Maccariello, Reyren, Garcia, Moutafis, Moreau-Luchaire, Collin, Bouzehouane,
  Cros, and Fert}}]{Legrand2017}
\bibinfo{author}{\bibfnamefont{W.}~\bibnamefont{Legrand}},
  \bibinfo{author}{\bibfnamefont{D.}~\bibnamefont{Maccariello}},
  \bibinfo{author}{\bibfnamefont{N.}~\bibnamefont{Reyren}},
  \bibinfo{author}{\bibfnamefont{K.}~\bibnamefont{Garcia}},
  \bibinfo{author}{\bibfnamefont{C.}~\bibnamefont{Moutafis}},
  \bibinfo{author}{\bibfnamefont{C.}~\bibnamefont{Moreau-Luchaire}},
  \bibinfo{author}{\bibfnamefont{S.}~\bibnamefont{Collin}},
  \bibinfo{author}{\bibfnamefont{K.}~\bibnamefont{Bouzehouane}},
  \bibinfo{author}{\bibfnamefont{V.}~\bibnamefont{Cros}}, \bibnamefont{and}
  \bibinfo{author}{\bibfnamefont{A.}~\bibnamefont{Fert}},
  \bibinfo{journal}{Nano Lett.} \textbf{\bibinfo{volume}{17}},
  \bibinfo{pages}{2703} (\bibinfo{year}{2017}).

\bibitem[{\citenamefont{Jiang et~al.}(2017)\citenamefont{Jiang, Zhang, Yu,
  Zhang, Wang, Benjamin~Jungfleisch, Pearson, Cheng, Heinonen, Wang
  et~al.}}]{Jiang2017}
\bibinfo{author}{\bibfnamefont{W.}~\bibnamefont{Jiang}},
  \bibinfo{author}{\bibfnamefont{X.}~\bibnamefont{Zhang}},
  \bibinfo{author}{\bibfnamefont{G.}~\bibnamefont{Yu}},
  \bibinfo{author}{\bibfnamefont{W.}~\bibnamefont{Zhang}},
  \bibinfo{author}{\bibfnamefont{X.}~\bibnamefont{Wang}},
  \bibinfo{author}{\bibfnamefont{M.}~\bibnamefont{Benjamin~Jungfleisch}},
  \bibinfo{author}{\bibfnamefont{J.~E.} \bibnamefont{Pearson}},
  \bibinfo{author}{\bibfnamefont{X.}~\bibnamefont{Cheng}},
  \bibinfo{author}{\bibfnamefont{O.}~\bibnamefont{Heinonen}},
  \bibinfo{author}{\bibfnamefont{K.~L.} \bibnamefont{Wang}},
  \bibnamefont{et~al.}, \bibinfo{journal}{Nat. Phys.}
  \textbf{\bibinfo{volume}{13}}, \bibinfo{pages}{162} (\bibinfo{year}{2017}).

\bibitem[{\citenamefont{Gross et~al.}(2018)\citenamefont{Gross, Akhtar, Hrabec,
  Sampaio, Mart\'{\i}nez, Chouaieb, Shields, Maletinsky, Thiaville, Rohart
  et~al.}}]{gross2018}
\bibinfo{author}{\bibfnamefont{I.}~\bibnamefont{Gross}},
  \bibinfo{author}{\bibfnamefont{W.}~\bibnamefont{Akhtar}},
  \bibinfo{author}{\bibfnamefont{A.}~\bibnamefont{Hrabec}},
  \bibinfo{author}{\bibfnamefont{J.}~\bibnamefont{Sampaio}},
  \bibinfo{author}{\bibfnamefont{L.~J.} \bibnamefont{Mart\'{\i}nez}},
  \bibinfo{author}{\bibfnamefont{S.}~\bibnamefont{Chouaieb}},
  \bibinfo{author}{\bibfnamefont{B.~J.} \bibnamefont{Shields}},
  \bibinfo{author}{\bibfnamefont{P.}~\bibnamefont{Maletinsky}},
  \bibinfo{author}{\bibfnamefont{A.}~\bibnamefont{Thiaville}},
  \bibinfo{author}{\bibfnamefont{S.}~\bibnamefont{Rohart}},
  \bibnamefont{et~al.}, \bibinfo{journal}{Phys. Rev. Materials}
  \textbf{\bibinfo{volume}{2}}, \bibinfo{pages}{024406} (\bibinfo{year}{2018}).

\bibitem[{\citenamefont{B\"uttner et~al.}(2017)\citenamefont{B\"uttner, Lemesh,
  Schneider, Pfau, G\"unther, Hessing, Geilhufe, Caretta, Engel, Kr\"uger
  et~al.}}]{buttner2017}
\bibinfo{author}{\bibfnamefont{F.}~\bibnamefont{B\"uttner}},
  \bibinfo{author}{\bibfnamefont{I.}~\bibnamefont{Lemesh}},
  \bibinfo{author}{\bibfnamefont{M.}~\bibnamefont{Schneider}},
  \bibinfo{author}{\bibfnamefont{B.}~\bibnamefont{Pfau}},
  \bibinfo{author}{\bibfnamefont{C.~M.} \bibnamefont{G\"unther}},
  \bibinfo{author}{\bibfnamefont{P.}~\bibnamefont{Hessing}},
  \bibinfo{author}{\bibfnamefont{J.}~\bibnamefont{Geilhufe}},
  \bibinfo{author}{\bibfnamefont{L.}~\bibnamefont{Caretta}},
  \bibinfo{author}{\bibfnamefont{D.}~\bibnamefont{Engel}},
  \bibinfo{author}{\bibfnamefont{B.}~\bibnamefont{Kr\"uger}},
  \bibnamefont{et~al.}, \bibinfo{journal}{Nat. Nanotech.}
  \textbf{\bibinfo{volume}{12}}, \bibinfo{pages}{1040} (\bibinfo{year}{2017}).

\bibitem[{\citenamefont{Woo et~al.}(2018{\natexlab{b}})\citenamefont{Woo, Song,
  Zhang, Ezawa, Zhou, Liu, Weigand, Finizio, Raabe, Park
  et~al.}}]{Woo2018nucle}
\bibinfo{author}{\bibfnamefont{S.}~\bibnamefont{Woo}},
  \bibinfo{author}{\bibfnamefont{K.~M.} \bibnamefont{Song}},
  \bibinfo{author}{\bibfnamefont{X.}~\bibnamefont{Zhang}},
  \bibinfo{author}{\bibfnamefont{M.}~\bibnamefont{Ezawa}},
  \bibinfo{author}{\bibfnamefont{Y.}~\bibnamefont{Zhou}},
  \bibinfo{author}{\bibfnamefont{X.}~\bibnamefont{Liu}},
  \bibinfo{author}{\bibfnamefont{M.}~\bibnamefont{Weigand}},
  \bibinfo{author}{\bibfnamefont{S.}~\bibnamefont{Finizio}},
  \bibinfo{author}{\bibfnamefont{J.}~\bibnamefont{Raabe}},
  \bibinfo{author}{\bibfnamefont{M.-C.} \bibnamefont{Park}},
  \bibnamefont{et~al.}, \bibinfo{journal}{Nature Electronics}
  \textbf{\bibinfo{volume}{1}}, \bibinfo{pages}{288}
  (\bibinfo{year}{2018}{\natexlab{b}}).

\bibitem[{\citenamefont{Mizukami et~al.}(2009)\citenamefont{Mizukami, Watanabe,
  Oogane, Ando, Miura, Shirai, and Miyazaki}}]{Mizukami2009}
\bibinfo{author}{\bibfnamefont{S.}~\bibnamefont{Mizukami}},
  \bibinfo{author}{\bibfnamefont{D.}~\bibnamefont{Watanabe}},
  \bibinfo{author}{\bibfnamefont{M.}~\bibnamefont{Oogane}},
  \bibinfo{author}{\bibfnamefont{Y.}~\bibnamefont{Ando}},
  \bibinfo{author}{\bibfnamefont{Y.}~\bibnamefont{Miura}},
  \bibinfo{author}{\bibfnamefont{M.}~\bibnamefont{Shirai}}, \bibnamefont{and}
  \bibinfo{author}{\bibfnamefont{T.}~\bibnamefont{Miyazaki}},
  \bibinfo{journal}{Journal of Applied Physics} \textbf{\bibinfo{volume}{105}},
  \bibinfo{pages}{07D306} (\bibinfo{year}{2009}).

\bibitem[{\citenamefont{Ludbrook et~al.}(2017)\citenamefont{Ludbrook, Dubuis,
  Puichaud, Ruck, and Granville}}]{Ludbrook2017}
\bibinfo{author}{\bibfnamefont{B.~M.} \bibnamefont{Ludbrook}},
  \bibinfo{author}{\bibfnamefont{G.}~\bibnamefont{Dubuis}},
  \bibinfo{author}{\bibfnamefont{A.-H.} \bibnamefont{Puichaud}},
  \bibinfo{author}{\bibfnamefont{B.~J.} \bibnamefont{Ruck}}, \bibnamefont{and}
  \bibinfo{author}{\bibfnamefont{S.}~\bibnamefont{Granville}},
  \bibinfo{journal}{Sci. Rep.} \textbf{\bibinfo{volume}{7}},
  \bibinfo{pages}{13620} (\bibinfo{year}{2017}).

\bibitem[{\citenamefont{Phatak et~al.}(2016)\citenamefont{Phatak, Heinonen,
  De~Graef, and Petford-Long}}]{phatak2016}
\bibinfo{author}{\bibfnamefont{C.}~\bibnamefont{Phatak}},
  \bibinfo{author}{\bibfnamefont{O.}~\bibnamefont{Heinonen}},
  \bibinfo{author}{\bibfnamefont{M.}~\bibnamefont{De~Graef}}, \bibnamefont{and}
  \bibinfo{author}{\bibfnamefont{A.}~\bibnamefont{Petford-Long}},
  \bibinfo{journal}{Nanolett.} \textbf{\bibinfo{volume}{16}},
  \bibinfo{pages}{4141} (\bibinfo{year}{2016}).

\bibitem[{\citenamefont{Rana et~al.}(2016)\citenamefont{Rana, Meshcheriakova,
  K\"ubler, Ernst, Karel, Hillebrand, Pippel, Werner, Nayak, and
  Felser}}]{rana2016}
\bibinfo{author}{\bibfnamefont{K.~G.} \bibnamefont{Rana}},
  \bibinfo{author}{\bibfnamefont{O.}~\bibnamefont{Meshcheriakova}},
  \bibinfo{author}{\bibfnamefont{J.}~\bibnamefont{K\"ubler}},
  \bibinfo{author}{\bibfnamefont{B.}~\bibnamefont{Ernst}},
  \bibinfo{author}{\bibfnamefont{J.}~\bibnamefont{Karel}},
  \bibinfo{author}{\bibfnamefont{R.}~\bibnamefont{Hillebrand}},
  \bibinfo{author}{\bibfnamefont{E.}~\bibnamefont{Pippel}},
  \bibinfo{author}{\bibfnamefont{P.}~\bibnamefont{Werner}},
  \bibinfo{author}{\bibfnamefont{A.}~\bibnamefont{Nayak}}, \bibnamefont{and}
  \bibinfo{author}{\bibfnamefont{C.}~\bibnamefont{Felser}},
  \bibinfo{journal}{New J. Phys.} \textbf{\bibinfo{volume}{18}},
  \bibinfo{pages}{085007} (\bibinfo{year}{2016}).

\bibitem[{\citenamefont{Rondin et~al.}(2014)\citenamefont{Rondin, Tetienne,
  Hingant, Roch, Maletinsky, and Jacques}}]{Rondin2014}
\bibinfo{author}{\bibfnamefont{L.}~\bibnamefont{Rondin}},
  \bibinfo{author}{\bibfnamefont{J.-P.} \bibnamefont{Tetienne}},
  \bibinfo{author}{\bibfnamefont{T.}~\bibnamefont{Hingant}},
  \bibinfo{author}{\bibfnamefont{J.-F.} \bibnamefont{Roch}},
  \bibinfo{author}{\bibfnamefont{P.}~\bibnamefont{Maletinsky}},
  \bibnamefont{and} \bibinfo{author}{\bibfnamefont{V.}~\bibnamefont{Jacques}},
  \bibinfo{journal}{Rep. Prog. Phys.} \textbf{\bibinfo{volume}{77}},
  \bibinfo{pages}{056503} (\bibinfo{year}{2014}).

\bibitem[{\citenamefont{Belmeguenai et~al.}(2018)\citenamefont{Belmeguenai,
  Roussign\'e, Bouloussa, Ch\'erif, Stashkevich, Nasui, Gabor,
  Mora-Hern\'andez, Nicholson, Inyang et~al.}}]{Belmeguenai2018}
\bibinfo{author}{\bibfnamefont{M.}~\bibnamefont{Belmeguenai}},
  \bibinfo{author}{\bibfnamefont{Y.}~\bibnamefont{Roussign\'e}},
  \bibinfo{author}{\bibfnamefont{H.}~\bibnamefont{Bouloussa}},
  \bibinfo{author}{\bibfnamefont{S.~M.} \bibnamefont{Ch\'erif}},
  \bibinfo{author}{\bibfnamefont{A.}~\bibnamefont{Stashkevich}},
  \bibinfo{author}{\bibfnamefont{M.}~\bibnamefont{Nasui}},
  \bibinfo{author}{\bibfnamefont{M.~S.} \bibnamefont{Gabor}},
  \bibinfo{author}{\bibfnamefont{A.}~\bibnamefont{Mora-Hern\'andez}},
  \bibinfo{author}{\bibfnamefont{B.}~\bibnamefont{Nicholson}},
  \bibinfo{author}{\bibfnamefont{O.-O.} \bibnamefont{Inyang}},
  \bibnamefont{et~al.}, \bibinfo{journal}{Phys. Rev. Applied}
  \textbf{\bibinfo{volume}{9}}, \bibinfo{pages}{044044} (\bibinfo{year}{2018}).

\bibitem[{\citenamefont{Di et~al.}(2015)\citenamefont{Di, Zhang, Lim, Ng, Kuok,
  Yu, Yoon, Qiu, and Yang}}]{KaiPRL2015}
\bibinfo{author}{\bibfnamefont{K.}~\bibnamefont{Di}},
  \bibinfo{author}{\bibfnamefont{V.~L.} \bibnamefont{Zhang}},
  \bibinfo{author}{\bibfnamefont{H.~S.} \bibnamefont{Lim}},
  \bibinfo{author}{\bibfnamefont{S.~C.} \bibnamefont{Ng}},
  \bibinfo{author}{\bibfnamefont{M.~H.} \bibnamefont{Kuok}},
  \bibinfo{author}{\bibfnamefont{J.}~\bibnamefont{Yu}},
  \bibinfo{author}{\bibfnamefont{J.}~\bibnamefont{Yoon}},
  \bibinfo{author}{\bibfnamefont{X.}~\bibnamefont{Qiu}}, \bibnamefont{and}
  \bibinfo{author}{\bibfnamefont{H.}~\bibnamefont{Yang}},
  \bibinfo{journal}{Phys. Rev. Lett.} \textbf{\bibinfo{volume}{114}},
  \bibinfo{pages}{047201} (\bibinfo{year}{2015}).

\bibitem[{\citenamefont{Belmeguenai et~al.}(2015)\citenamefont{Belmeguenai,
  Adam, Roussign\'e, Eimer, Devolder, Kim, Cherif, Stashkevich, and
  Thiaville}}]{Belmeguenai2015}
\bibinfo{author}{\bibfnamefont{M.}~\bibnamefont{Belmeguenai}},
  \bibinfo{author}{\bibfnamefont{J.-P.} \bibnamefont{Adam}},
  \bibinfo{author}{\bibfnamefont{Y.}~\bibnamefont{Roussign\'e}},
  \bibinfo{author}{\bibfnamefont{S.}~\bibnamefont{Eimer}},
  \bibinfo{author}{\bibfnamefont{T.}~\bibnamefont{Devolder}},
  \bibinfo{author}{\bibfnamefont{J.-V.} \bibnamefont{Kim}},
  \bibinfo{author}{\bibfnamefont{S.~M.} \bibnamefont{Cherif}},
  \bibinfo{author}{\bibfnamefont{A.}~\bibnamefont{Stashkevich}},
  \bibnamefont{and}
  \bibinfo{author}{\bibfnamefont{A.}~\bibnamefont{Thiaville}},
  \bibinfo{journal}{Phys. Rev. B} \textbf{\bibinfo{volume}{91}},
  \bibinfo{pages}{180405} (\bibinfo{year}{2015}).

\bibitem[{\citenamefont{Maletinsky et~al.}(2012)\citenamefont{Maletinsky, Hong,
  Grinolds, Hausmann, Lukin, Walsworth, Lon\u{c}ar, and
  Yacoby}}]{Maletinsky2012}
\bibinfo{author}{\bibfnamefont{P.}~\bibnamefont{Maletinsky}},
  \bibinfo{author}{\bibfnamefont{S.}~\bibnamefont{Hong}},
  \bibinfo{author}{\bibfnamefont{M.~S.} \bibnamefont{Grinolds}},
  \bibinfo{author}{\bibfnamefont{B.}~\bibnamefont{Hausmann}},
  \bibinfo{author}{\bibfnamefont{M.~D.} \bibnamefont{Lukin}},
  \bibinfo{author}{\bibfnamefont{R.~L.} \bibnamefont{Walsworth}},
  \bibinfo{author}{\bibfnamefont{M.}~\bibnamefont{Lon\u{c}ar}},
  \bibnamefont{and} \bibinfo{author}{\bibfnamefont{A.}~\bibnamefont{Yacoby}},
  \bibinfo{journal}{Nat. Nanotech.} \textbf{\bibinfo{volume}{7}},
  \bibinfo{pages}{320} (\bibinfo{year}{2012}).

\bibitem[{\citenamefont{Appel et~al.}(2016)\citenamefont{Appel, Neu, Ganzhorn,
  Barfuss, Batzer, Gratz, Tschope, and Maletinsky}}]{Appel2016}
\bibinfo{author}{\bibfnamefont{P.}~\bibnamefont{Appel}},
  \bibinfo{author}{\bibfnamefont{E.}~\bibnamefont{Neu}},
  \bibinfo{author}{\bibfnamefont{M.}~\bibnamefont{Ganzhorn}},
  \bibinfo{author}{\bibfnamefont{A.}~\bibnamefont{Barfuss}},
  \bibinfo{author}{\bibfnamefont{M.}~\bibnamefont{Batzer}},
  \bibinfo{author}{\bibfnamefont{M.}~\bibnamefont{Gratz}},
  \bibinfo{author}{\bibfnamefont{A.}~\bibnamefont{Tschope}}, \bibnamefont{and}
  \bibinfo{author}{\bibfnamefont{P.}~\bibnamefont{Maletinsky}},
  \bibinfo{journal}{Review of Scientific Instruments}
  \textbf{\bibinfo{volume}{87}}, \bibinfo{pages}{063703}
  (\bibinfo{year}{2016}).

\bibitem[{\citenamefont{Epstein et~al.}(2005)\citenamefont{Epstein, Mendoza,
  Kato, and Awschalom}}]{Epstein2005}
\bibinfo{author}{\bibfnamefont{R.~J.} \bibnamefont{Epstein}},
  \bibinfo{author}{\bibfnamefont{F.~M.} \bibnamefont{Mendoza}},
  \bibinfo{author}{\bibfnamefont{Y.~K.} \bibnamefont{Kato}}, \bibnamefont{and}
  \bibinfo{author}{\bibfnamefont{D.~D.} \bibnamefont{Awschalom}},
  \bibinfo{journal}{Nat. Phys} \textbf{\bibinfo{volume}{1}},
  \bibinfo{pages}{94} (\bibinfo{year}{2005}).

\bibitem[{\citenamefont{Lai et~al.}(2009)\citenamefont{Lai, Zheng, Jelezko,
  Treussart, and Roch}}]{Lai2009}
\bibinfo{author}{\bibfnamefont{N.}~\bibnamefont{Lai}},
  \bibinfo{author}{\bibfnamefont{D.}~\bibnamefont{Zheng}},
  \bibinfo{author}{\bibfnamefont{F.}~\bibnamefont{Jelezko}},
  \bibinfo{author}{\bibfnamefont{F.}~\bibnamefont{Treussart}},
  \bibnamefont{and} \bibinfo{author}{\bibfnamefont{J.-F.} \bibnamefont{Roch}},
  \bibinfo{journal}{Appl. Phys. Lett.} \textbf{\bibinfo{volume}{95}},
  \bibinfo{pages}{133101} (\bibinfo{year}{2009}).

\bibitem[{\citenamefont{Rondin et~al.}(2012)\citenamefont{Rondin, Tetienne,
  Spinicelli, Dal~Savio, Karrai, Dantelle, Thiaville, Rohart, Roch, and
  Jacques}}]{Rondin2012}
\bibinfo{author}{\bibfnamefont{L.}~\bibnamefont{Rondin}},
  \bibinfo{author}{\bibfnamefont{J.-P.} \bibnamefont{Tetienne}},
  \bibinfo{author}{\bibfnamefont{P.}~\bibnamefont{Spinicelli}},
  \bibinfo{author}{\bibfnamefont{C.}~\bibnamefont{Dal~Savio}},
  \bibinfo{author}{\bibfnamefont{K.}~\bibnamefont{Karrai}},
  \bibinfo{author}{\bibfnamefont{G.}~\bibnamefont{Dantelle}},
  \bibinfo{author}{\bibfnamefont{A.}~\bibnamefont{Thiaville}},
  \bibinfo{author}{\bibfnamefont{S.}~\bibnamefont{Rohart}},
  \bibinfo{author}{\bibfnamefont{J.-F.} \bibnamefont{Roch}}, \bibnamefont{and}
  \bibinfo{author}{\bibfnamefont{V.}~\bibnamefont{Jacques}},
  \bibinfo{journal}{Appl. Phys. Lett.} \textbf{\bibinfo{volume}{100}},
  \bibinfo{eid}{153118} (\bibinfo{year}{2012}).

\bibitem[{\citenamefont{Tetienne et~al.}(2012)\citenamefont{Tetienne, Rondin,
  Spinicelli, Chipaux, Debuisschert, Roch, and Jacques}}]{Tetienne2012}
\bibinfo{author}{\bibfnamefont{J.-P.} \bibnamefont{Tetienne}},
  \bibinfo{author}{\bibfnamefont{L.}~\bibnamefont{Rondin}},
  \bibinfo{author}{\bibfnamefont{P.}~\bibnamefont{Spinicelli}},
  \bibinfo{author}{\bibfnamefont{M.}~\bibnamefont{Chipaux}},
  \bibinfo{author}{\bibfnamefont{T.}~\bibnamefont{Debuisschert}},
  \bibinfo{author}{\bibfnamefont{J.-F.} \bibnamefont{Roch}}, \bibnamefont{and}
  \bibinfo{author}{\bibfnamefont{V.}~\bibnamefont{Jacques}},
  \bibinfo{journal}{New J. Phys.} \textbf{\bibinfo{volume}{14}},
  \bibinfo{pages}{103033} (\bibinfo{year}{2012}).

\bibitem[{\citenamefont{Hingant et~al.}(2015)\citenamefont{Hingant, Tetienne,
  Mart\'{\i}nez, Garcia, Ravelosona, Roch, and Jacques}}]{Hingant2015}
\bibinfo{author}{\bibfnamefont{T.}~\bibnamefont{Hingant}},
  \bibinfo{author}{\bibfnamefont{J.-P.} \bibnamefont{Tetienne}},
  \bibinfo{author}{\bibfnamefont{L.~J.} \bibnamefont{Mart\'{\i}nez}},
  \bibinfo{author}{\bibfnamefont{K.}~\bibnamefont{Garcia}},
  \bibinfo{author}{\bibfnamefont{D.}~\bibnamefont{Ravelosona}},
  \bibinfo{author}{\bibfnamefont{J.-F.} \bibnamefont{Roch}}, \bibnamefont{and}
  \bibinfo{author}{\bibfnamefont{V.}~\bibnamefont{Jacques}},
  \bibinfo{journal}{Phys. Rev. Applied} \textbf{\bibinfo{volume}{4}},
  \bibinfo{pages}{014003} (\bibinfo{year}{2015}).

\bibitem[{\citenamefont{Zeissler et~al.}(2017)\citenamefont{Zeissler,
  Mruczkiewicz, Finizio, Raabe, Shepley, Sadovnikov, Nikitov, Fallon,
  McFadzean, McVitie et~al.}}]{Zeissler2017}
\bibinfo{author}{\bibfnamefont{K.}~\bibnamefont{Zeissler}},
  \bibinfo{author}{\bibfnamefont{M.}~\bibnamefont{Mruczkiewicz}},
  \bibinfo{author}{\bibfnamefont{S.}~\bibnamefont{Finizio}},
  \bibinfo{author}{\bibfnamefont{J.}~\bibnamefont{Raabe}},
  \bibinfo{author}{\bibfnamefont{P.~M.} \bibnamefont{Shepley}},
  \bibinfo{author}{\bibfnamefont{A.~V.} \bibnamefont{Sadovnikov}},
  \bibinfo{author}{\bibfnamefont{S.~A.} \bibnamefont{Nikitov}},
  \bibinfo{author}{\bibfnamefont{K.}~\bibnamefont{Fallon}},
  \bibinfo{author}{\bibfnamefont{S.}~\bibnamefont{McFadzean}},
  \bibinfo{author}{\bibfnamefont{S.}~\bibnamefont{McVitie}},
  \bibnamefont{et~al.}, \bibinfo{journal}{Sci. Rep.}
  \textbf{\bibinfo{volume}{7}}, \bibinfo{pages}{15125} (\bibinfo{year}{2017}).

\bibitem[{\citenamefont{Juge et~al.}(2018)\citenamefont{Juge, Je,
  de~Souza~Chaves, Pizzini, Buda-Prejbeanu, Aballe, Foerster, Locatelli,
  Onur~Mentes, Sala et~al.}}]{Juge2018}
\bibinfo{author}{\bibfnamefont{R.}~\bibnamefont{Juge}},
  \bibinfo{author}{\bibfnamefont{S.-G.} \bibnamefont{Je}},
  \bibinfo{author}{\bibfnamefont{D.}~\bibnamefont{de~Souza~Chaves}},
  \bibinfo{author}{\bibfnamefont{S.}~\bibnamefont{Pizzini}},
  \bibinfo{author}{\bibfnamefont{L.~D.} \bibnamefont{Buda-Prejbeanu}},
  \bibinfo{author}{\bibfnamefont{L.}~\bibnamefont{Aballe}},
  \bibinfo{author}{\bibfnamefont{M.}~\bibnamefont{Foerster}},
  \bibinfo{author}{\bibfnamefont{A.}~\bibnamefont{Locatelli}},
  \bibinfo{author}{\bibfnamefont{T.}~\bibnamefont{Onur~Mentes}},
  \bibinfo{author}{\bibfnamefont{A.}~\bibnamefont{Sala}}, \bibnamefont{et~al.},
  \bibinfo{journal}{J. Magn. Magn. Mater.} \textbf{\bibinfo{volume}{455}},
  \bibinfo{pages}{3 } (\bibinfo{year}{2018}).

\bibitem[{\citenamefont{Lo~Conte et~al.}(2014)\citenamefont{Lo~Conte, Hrabec,
  Mihai, Schulz, Noh, Marrows, Moore, and Kl\"aui}}]{loconte2014}
\bibinfo{author}{\bibfnamefont{R.}~\bibnamefont{Lo~Conte}},
  \bibinfo{author}{\bibfnamefont{A.}~\bibnamefont{Hrabec}},
  \bibinfo{author}{\bibfnamefont{A.}~\bibnamefont{Mihai}},
  \bibinfo{author}{\bibfnamefont{T.}~\bibnamefont{Schulz}},
  \bibinfo{author}{\bibfnamefont{S.-J.} \bibnamefont{Noh}},
  \bibinfo{author}{\bibfnamefont{C.~H.} \bibnamefont{Marrows}},
  \bibinfo{author}{\bibfnamefont{T.~A.} \bibnamefont{Moore}}, \bibnamefont{and}
  \bibinfo{author}{\bibfnamefont{M.}~\bibnamefont{Kl\"aui}},
  \bibinfo{journal}{Appl. Phys. Lett.} \textbf{\bibinfo{volume}{105}},
  \bibinfo{pages}{122404} (\bibinfo{year}{2014}).

\bibitem[{\citenamefont{Liu et~al.}(2012)\citenamefont{Liu, Lee, Gudmundsen,
  Ralph, and Buhrman}}]{liu2012}
\bibinfo{author}{\bibfnamefont{L.}~\bibnamefont{Liu}},
  \bibinfo{author}{\bibfnamefont{O.~J.} \bibnamefont{Lee}},
  \bibinfo{author}{\bibfnamefont{T.~J.} \bibnamefont{Gudmundsen}},
  \bibinfo{author}{\bibfnamefont{D.~C.} \bibnamefont{Ralph}}, \bibnamefont{and}
  \bibinfo{author}{\bibfnamefont{R.~A.} \bibnamefont{Buhrman}},
  \bibinfo{journal}{Phys. Rev. Lett.} \textbf{\bibinfo{volume}{109}},
  \bibinfo{pages}{096602} (\bibinfo{year}{2012}).

\bibitem[{\citenamefont{Vansteenkiste et~al.}(2014)\citenamefont{Vansteenkiste,
  Leliaert, Dvornik, Helsen, Garcia-Sanchez, and Van~Waeyenberge}}]{MuMax}
\bibinfo{author}{\bibfnamefont{A.}~\bibnamefont{Vansteenkiste}},
  \bibinfo{author}{\bibfnamefont{J.}~\bibnamefont{Leliaert}},
  \bibinfo{author}{\bibfnamefont{M.}~\bibnamefont{Dvornik}},
  \bibinfo{author}{\bibfnamefont{M.}~\bibnamefont{Helsen}},
  \bibinfo{author}{\bibfnamefont{F.}~\bibnamefont{Garcia-Sanchez}},
  \bibnamefont{and}
  \bibinfo{author}{\bibfnamefont{B.}~\bibnamefont{Van~Waeyenberge}},
  \bibinfo{journal}{AIP Advances} \textbf{\bibinfo{volume}{4}},
  \bibinfo{pages}{107133} (\bibinfo{year}{2014}).

\bibitem[{\citenamefont{Emori et~al.}(2013)\citenamefont{Emori, Bauer, Ahn,
  Martinez, and Beach}}]{Emori2013}
\bibinfo{author}{\bibfnamefont{S.}~\bibnamefont{Emori}},
  \bibinfo{author}{\bibfnamefont{U.}~\bibnamefont{Bauer}},
  \bibinfo{author}{\bibfnamefont{S.-M.} \bibnamefont{Ahn}},
  \bibinfo{author}{\bibfnamefont{E.}~\bibnamefont{Martinez}}, \bibnamefont{and}
  \bibinfo{author}{\bibfnamefont{G.~S.~D.} \bibnamefont{Beach}},
  \bibinfo{journal}{Nat. Mater.} \textbf{\bibinfo{volume}{12}},
  \bibinfo{pages}{611} (\bibinfo{year}{2013}).

\bibitem[{\citenamefont{Jiang et~al.}(2016)\citenamefont{Jiang, Zhang, Yu,
  Zhang, Wang, Jungfleisch, Pearson, Cheng, Heinonen, Wang et~al.}}]{Jiang2016}
\bibinfo{author}{\bibfnamefont{W.}~\bibnamefont{Jiang}},
  \bibinfo{author}{\bibfnamefont{X.}~\bibnamefont{Zhang}},
  \bibinfo{author}{\bibfnamefont{G.}~\bibnamefont{Yu}},
  \bibinfo{author}{\bibfnamefont{W.}~\bibnamefont{Zhang}},
  \bibinfo{author}{\bibfnamefont{X.}~\bibnamefont{Wang}},
  \bibinfo{author}{\bibfnamefont{M.~B.} \bibnamefont{Jungfleisch}},
  \bibinfo{author}{\bibfnamefont{J.~E.} \bibnamefont{Pearson}},
  \bibinfo{author}{\bibfnamefont{X.}~\bibnamefont{Cheng}},
  \bibinfo{author}{\bibfnamefont{O.}~\bibnamefont{Heinonen}},
  \bibinfo{author}{\bibfnamefont{K.~L.} \bibnamefont{Wang}},
  \bibnamefont{et~al.}, \bibinfo{journal}{Nat. Phys.}
  \textbf{\bibinfo{volume}{13}}, \bibinfo{pages}{162} (\bibinfo{year}{2016}).

\end{thebibliography}

\end{document}